\documentclass[lineno]{jfm}

\usepackage{graphicx}
\usepackage{newtxtext}
\usepackage{newtxmath}
\usepackage{natbib}
\usepackage{hyperref}
\usepackage{caption}
\hypersetup{
    colorlinks = true,
    urlcolor   = blue,
    citecolor  = black,
}

\newcommand{\RomanNumeralCaps}[1]

\newcommand{\la}{\left<}
\newcommand{\ra}{\right>}

\newcommand{\cor}[1]{{#1}}

\shorttitle{Selection of minimum enstrophy solutions}
\shortauthor{B. Gallet}

\title{Two-dimensional turbulence above topography: condensation transition and selection of minimum enstrophy solutions}

\author{Basile Gallet\aff{1}
  \corresp{\email{basile.gallet@cea.fr}}}

\affiliation{\aff{1}{Universit\'e Paris-Saclay, CNRS, CEA, Service de Physique de l'Etat Condens\'e, 91191 Gif-sur-Yvette, France.}}

\begin{document}
\maketitle


\begin{abstract}
We consider two-dimensional flows above topography, revisiting the selective decay (or minimum-enstrophy) hypothesis of Bretherton and Haidvogel. We derive a `condensed branch' of solutions to the variational problem where a domain-scale condensate coexists with a flow at the (smaller) scale of the topography. The condensate arises through a supercritical bifurcation as the conserved energy of the initial condition exceeds a threshold value, a prediction that we quantitatively validate using Direct Numerical Simulations (DNS). We then consider the forced-dissipative case, showing how weak forcing and dissipation select a single dissipative state out of the continuum of solutions to the energy-conserving system predicted by selective decay. As the forcing strength increases, the condensate arises through a supercritical bifurcation for topographic-scale forcing and through a subcritical bifurcation for domain-scale forcing, both predictions being quantitatively validated by DNS. This method provides a way of determining the equilibrated state of forced-dissipative flows based on variational approaches to the associated energy-conserving system, such as the statistical mechanics of 2D flows or selective decay.
\end{abstract}



\section{Introduction}

Rapidly rotating flows above topography are both a standard model for various large-scale oceanic flows and a fascinating playground to study how disorder disrupts the large-scale self-organization of two-dimensional flows. A first oceanic situation of interest is the Antarctic Circumpolar Current (ACC), where turbulent eddies are channeled by and somewhat locked to the topography, with important consequences for both the turbulent transport and the speed of the ACC itself~\citep{straub1993transport,nadeau2009basin,abernathey2014topographic,nadeau2015role,constantinou2017beta,constantinou2018barotropic,constantinou2019eddy,barthel2022baroclinic,yung2022topographic}. At a more local scale, the Lofoten vortex has attracted significant attention~\citep{kohl2007generation,soiland2013structure,trodahl2020regeneration,lacasce2024vortices} because its direction of rotation is inconsistent with the natural tendency for flows to homogenize potential vorticity~\citep{rhines1982homogenization}.

Early on, the study of rotating flows above topography has been connected to the search for general organizing principles governing the evolution of quasi-two-dimensional flows. \citet{bretherton1976two} (BH in the following) applied `selective decay' to such flows, an organizing principle borrowed from plasma physics~\citep{kruskal1958equilibrium,taylor1974relaxation,montgomery1978three}. According to the selective decay principle, the forward-cascading invariant should be minimized for a given (conserved) value of the inverse-cascading invariant. Based on the inverse energy cascade and forward enstrophy cascade of 2D flows, BH thus conjectured that the system evolves towards a state of minimum enstrophy, while conserving its initial energy. They report reasonable, although imperfect, agreement with their numerical simulations.
A second, more recent variational approach is the statistical mechanics of two-dimensional flows~\citep{miller1990statistical,robert1991maximum,robert1991statistical,bouchet2012statistical}. The approach has been applied to several two dimensional fluid systems, predicting various branches of solutions with different flow structures~\citep{,naso2011statistical}, together with `phase transitions' between these various states~\citep{chavanis1996classification,venaille2009statistical}. In its most general form, the theory is based on the conservation of all the invariants of the system: energy, all the moments of the vorticity field, together with linear momentum, angular momentum or circulation, depending on the geometry of interest. However, for most practical computations the high-order (potential-)vorticity moments are discarded~(e.g. \citet{salmon1976equilibrium,venaille2012bottom}) and the theory ends up being equivalent to selective decay~\citep{carnevale1987nonlinear}, namely enstrophy minimization subject to the conservation of some first and second-order invariants (see \citet{brands1999maximum} and \citet{nycander2004stable} for examples where all the invariants are simultaneously considered). While they provide invaluable insight, an important limitation of such variational approaches is that they apply to energy-conserving systems only, as opposed to forced-dissipative ones. Bridging this gap is a central motivation for the present study.


Recently, \citet{siegelman2023two} (SY in the following) revisited two-dimensional turbulence above topography, with a focus on the difference between the low-energy flows and the high-energy flows.  In qualitative agreement with selective decay, enstrophy robustly decreases with time in their numerical simulations, while energy is conserved to a very good approximation. However, SY also notice that the flows attained in the long-time limit depart from the predictions of selective decay. For low initial energy, the numerical solutions exhibit intense cyclones above topographic mounts and anticyclones above topographic depressions that are not predicted by selective decay. \cor{Instead, such pinned vortices result from the tendency for anticyclones (cyclones) to migrate down-slope (up-slope), resulting in a vortex segregation mechanism that opposes PV homogenization~\citep{carnevale1991propagation,kohl2007generation,trodahl2020regeneration,solodoch2021formation}.}
Above a threshold value of the initial energy, these vortices are no longer pinned to the topography, and instead they roam within the numerical domain.

In the present study, we consider a spatial structure for the topography that differs from both BH and SY. While these authors include topography at all scales, we restrict attention to topography at smaller scale than the extent of the fluid domain (strictly, although no scale separation is needed). Our goal is threefold. First, we wish to continue the program initiated by SY and further characterize the transition from the low energy states to the high energy states: how does the flow transition from a topographically-locked state to the typical large-scale condensates that characterize weakly damped two-dimensional flows~\citep{kraichnan1975statistical,gallet2013two,laurie2014universal}? Secondly, we want to assess the predictive skill of the aforementioned variational approaches: do they predict a sharp bifurcation or a smooth crossover between the two types of flows? Is the transition continuous or discontinuous? Can the theory quantitatively predict the strength of the large-scale condensate? Thirdly, we want to address the limitation of the variational approaches mentioned above, namely that they were designed for energy-conserving systems only, whereas natural and laboratory flows are typically forced and dissipative. While the conserved invariants of the system are the natural control parameters of the variational theories, such quantities are no longer conserved in the presence of forcing and dissipation. How to make quantitative predictions in this context? How to leverage the variational approaches to predict the equilibrated state of the forced-dissipative system?


We introduce the theoretical setup in section~\ref{sec:Rapidly}, before focusing on the energy-conserving system in section~\ref{sec:Energy}. We predict a transition to condensation based on selective decay. In section~\ref{sec:Numerical}, we report numerical simulations that quantitatively validate the predicted amplitude of the large-scale condensate. We turn to the forced-dissipative system in section~\ref{sec:Forced}. We show how weak forcing and damping select a single state out of the continuum of solutions to the energy-conserving system predicted by selective decay. We compare the predicted amplitude of the condensate to Direct Numerical Simulations (DNS) of the forced dissipative system, obtaining very good agreement for both topographic-scale forcing and domain-scale forcing. We conclude in section~\ref{sec:Conclusion}.

\section{Rapidly rotating flows above topography\label{sec:Rapidly}}

We consider a shallow-water flow above weak topography in a rapidly rotating frame, described within the \cor{$f$-plane} quasi-geostrophic approximation. The horizontal velocity field ${\bf u}(x,y,t)$ stems from a streamfunction, ${\bf u} = - \bnabla \times [\psi(x,y,t) {\bf e}_z]$. The evolution of the system is governed by the material conservation of potential vorticity (PV) $q(x,y,t)$:
\begin{eqnarray}
& & \partial_t q + J(\psi,q) = {\cal F}(x,y)+{\cal D} \, , \label{eqq}\\
& & q= \Delta \psi + \eta(x,y) \, , \label{defq}
\end{eqnarray}
where \cor{$J(f,g)=(\partial_x f) (\partial_y g) - (\partial_y f)(\partial_x g)$ denotes the Jacobian operator, $\Delta$ denotes the Laplacian operator} and $\eta(x,y)$ denotes the `topographic' PV. For a layer of fluid of local depth $H+h(x,y)$, where $h(x,y) \ll H$ denotes the mean-zero fluctuations in bathymetry, the topographic PV is $\eta(x,y)=-f_0 \, h(x,y)/H$, with $f_0$ the Coriolis parameter. We consider equations (\ref{eqq}-\ref{defq}) inside a doubly periodic domain, $(x,y)\in [0,2\pi L]^2$, with $2\pi L$ greater than the greatest wavelength of the topography (strictly, although no scale separation is required). 
We include forcing and dissipation on the right-hand side of equation (\ref{eqq}). The forcing term ${\cal F}(x,y)$ corresponds to the wind-stress curl in a typical ocean setting~\citep{Salmonbook}. The dissipative term consists of linear `Ekman' bottom drag\cor{~\citep{Pedloskybook}} and hyperviscosity, ${\cal D}=-\kappa \Delta \psi - \nu \Delta^3 \psi$. In the following we consider both the `energy-conserving  case', characterized by ${\cal F}=0$, $\kappa=0$ and small hyperviscosity $\nu$, and the `forced dissipative case', characterized by ${\cal F} \neq 0$ and $\kappa \neq 0$.

\section{Energy-conserving case: transition to condensation\label{sec:Energy}}

When the right-hand side of (\ref{eqq}) is set to zero, $ {\cal F}= {\cal D}=0$, the system possesses two quadratic invariants. Denoting space average as $\la \cdot \ra$, the first invariant is the kinetic energy $E=\la |\bnabla \psi|^2 \ra$, where we omit the prefactor $1/2$ to alleviate the algebra. The second invariant is the enstrophy $Q=\la q^2 \ra$, omitting again the prefactor $1/2$. In a similar fashion to two-dimensional turbulence in the absence of topography, one expects the transient dynamics of the system to be characterized by an inverse cascade of energy together with a forward cascade of enstrophy. If a small hyperviscosity $\nu \neq 0$ is included, we expect $Q$ to be robustly dissipated  \cor{once enstrophy has reached the small viscous dissipative scale}. \cor{The Kolmogorov-Kraichnan dual cascade phenomenology further suggests that such `anomalous'} dissipation of enstrophy is maintained as $\nu \to 0$, whereas the energy $E$ is conserved in that limit. That enstrophy is strongly dissipated by weak small-scale damping while energy is approximately conserved is clearly illustrated by the recent DNS of SY.
This intuition led BH to conjecture that the system eventually reaches the minimum-enstrophy state associated with the conserved energy of the initial condition. This state is obtained by extremalizing the functional ${\cal L}\{ \psi \}=Q+\mu E$, where $\mu$ is a Lagrange multiplier ensuring energy conservation. \cor{This variational problem} leads to:
\begin{eqnarray}
q = \Delta \psi + \eta = \mu \psi \, . \label{MEH}
\end{eqnarray}
Note that the same relation can be deduced from the statistical mechanics of 2D flows, if one retains only the invariants $E$ and $Q$. An argument in favor of selective decay as opposed to the statistical mechanics of conservative systems is that $Q$ consistently decreases during the transient phase observed in DNS.

An appealing aspect of enstrophy minimization is that fields satisfying (\ref{MEH}) are steady solutions to equation (\ref{eqq}) with ${\cal F}={\cal D}=0$. Additionally, the solution to (\ref{MEH}) corresponding to the absolute enstrophy minimum is stable according to Arnol'd's stability criterion~\citep{arnol2013mathematical}. The following subsections aim at computing this absolute minimizer.

\subsection{The Bretherton-Haidvogel branch}

The standard approach to solving equation (\ref{MEH}) is the one proposed by BH. They assume that $\mu>-k_0^2$, where $k_0=1/L$ denotes the minimum (gravest) wavenumber of the square domain. \cor{Introducing the Fourier decompositions of $\eta$ and $\psi$ as:
\begin{eqnarray}
\eta(x,y) = \sum_{{\bf k}L \in \mathbb{Z}^2} \hat{\eta}_{\bf k} e^{i {\bf k} \cdot \bf{x}}  \, , \qquad \psi(x,y,t) = \sum_{{\bf k}L \in \mathbb{Z}^2} \hat{\psi}_{\bf k}(t) e^{i {\bf k} \cdot \bf{x}} \, , \label{eq:Fourier}
\end{eqnarray}}
equation (\ref{MEH}) has a unique solution obtained in spectral space:
\begin{eqnarray}
\hat{\psi}_{\bf k} =\frac{\hat{\eta}_{\bf k}}{\mu+k^2} \, .\label{psiBH}
\end{eqnarray}
From (\ref{psiBH}) the energy $E$ and the enstrophy $Q$ can be expressed in terms of $\mu$. This leads to a parametric curve for $Q$ versus $E$, and one thus predicts (among other things) the final enstrophy in terms of the initial energy of the system.

\subsection{Monoscale topography}

Although not a necessary assumption, restricting attention to monoscale topography provides an illustrative example. We thus consider that the topography is peaked around a single wavenumber $k_\eta$, such that $\Delta \eta = - k_\eta^2 \eta$. For such topography (\ref{psiBH}) leads to:
\begin{eqnarray}
\psi  =  \frac{\eta}{\mu+k_\eta^2} \, , \qquad E  =  \frac{k_\eta^2}{(\mu+k_\eta^2)^2} \la \eta^2 \ra \, , \qquad Q  = \frac{\mu^2}{(\mu+k_\eta^2)^2}  \la \eta^2 \ra \, .
\end{eqnarray}
\cor{It proves convenient to introduce the energy scale $E_\eta=\la \eta^2 \ra/k_\eta^2$ associated with the topography (following SY, $E_\eta$ also corresponds to the energy of a flow with homogenized PV, see discussion in section~\ref{sec:Conclusion}). }
\cor{Denoting the dimensionless energy as $\tilde{E}=E/E_\eta$, one can eliminate $\mu$ to express $Q$ in terms of~$E$:}
\begin{eqnarray}
\frac{Q}{\la \eta^2 \ra} & = & \left(\sqrt{\tilde{E}}-1 \right)^2 \, . \label{QBH}
\end{eqnarray}

\subsection{The condensed branch}

\begin{figure}
    \centerline{\includegraphics[width=9 cm]{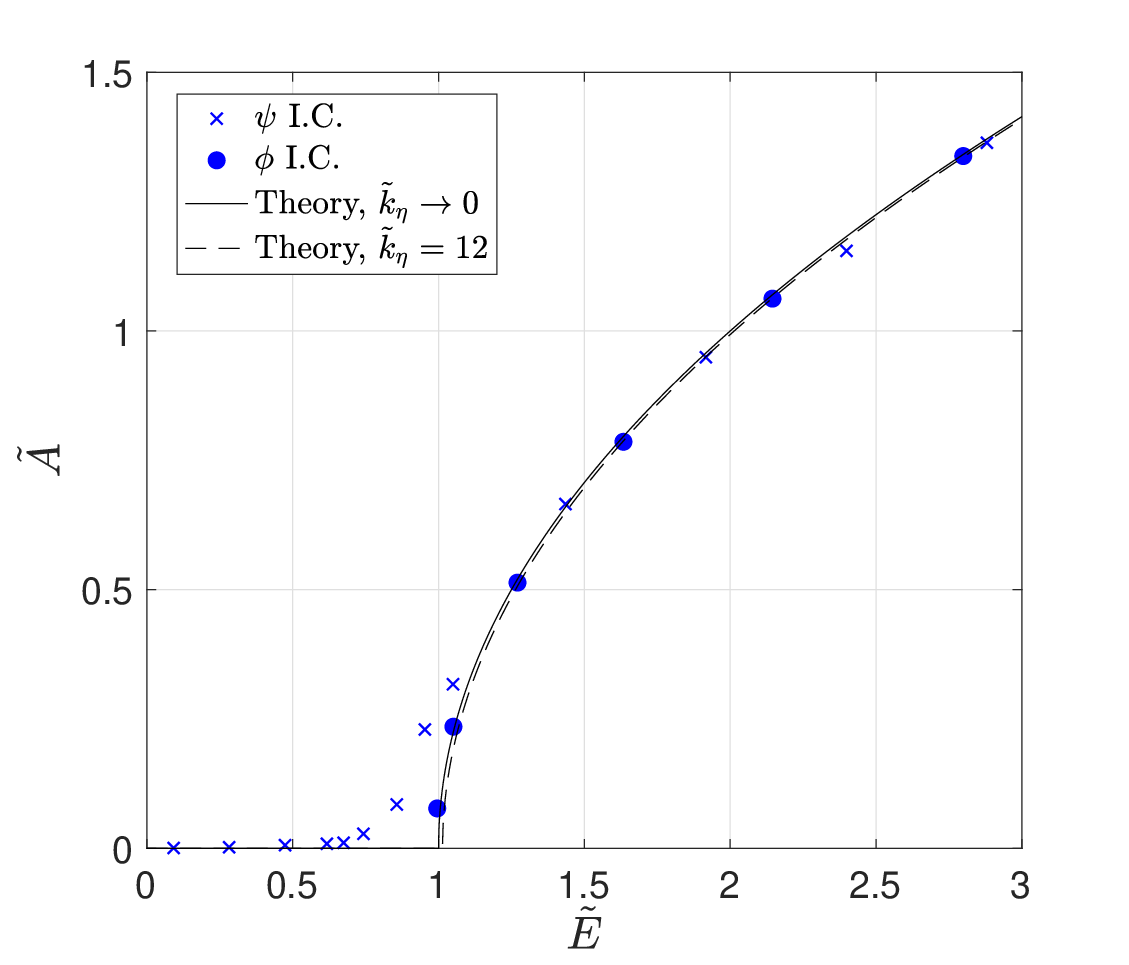} }
   \caption{The amplitude $\tilde{A}$ of the condensate as a function of the (conserved) energy of the initial condition $\tilde{E}$. The theory predicts the existence of a condensed branch for $\tilde{E}>1$ in the scale separation limit (slightly larger threshold for finite $\tilde{k}_\eta$). \cor{The DNS are performed using $\tilde{k}_\eta=12$. The data points} agree well with the predicted amplitude, for the two initial conditions introduced in section \ref{sec:Numerical}. Some departure from the theory remains in the immediate vicinity of the threshold for condensation, as a consequence of the extra vortices pinned to the topography.\label{fig:AvsE}}
\end{figure}

When the system has a size greater than the largest scale of the topography (when $k_0=1/L<k_\eta$ for monoscale topography) we argue that the BH branch above is not always the absolute minimizer of the enstrophy. Indeed, consider equation (\ref{MEH}) with $\mu=-k_0^2$:
\begin{eqnarray}
\Delta \psi + k_0^2 \psi = - \eta \, . 
\end{eqnarray}
The operator on the left-hand side is not invertible anymore and the solution is (restricting attention to monoscale topography, the generalization to an arbitrary spectrum for the topography being straightforward):
\begin{eqnarray}
\psi(x,y)  & = &  \psi_\eta(x,y) + A \, \psi_0(x,y) \, , \label{eq:psicondensed} \\ 
\psi_\eta(x,y) & = & \frac{\eta(x,y)}{k_\eta^2-k_0^2} \, .
\end{eqnarray}
\cor{In equation (\ref{eq:psicondensed}),} $\psi_0(x,y)$ is a harmonic function at the gravest scale of the domain (that is, $\Delta \psi_0 =- k_0^2 \psi_0$) normalized such that $\la \psi_0^2 \ra=1$\cor{, and $A \geq 0$ denotes the amplitude of this large-scale `condensate'}. 
\cor{$A$ and $\psi_0(x,y)$ are related to the two gravest Fourier coefficients of $\psi$ through $A \, \psi_0(x,y) = \hat{\psi}_{(1/L,0)} e^{i x/L} + \hat{\psi}_{(0,1/L)} e^{i y/L}+ \text{c.c.}$, that is:
\begin{eqnarray}
A =\sqrt{2 |\hat{\psi}_{(1/L,0)}|^2 + 2 |\hat{\psi}_{(0,1/L)}|^2} \, . \label{eq:Adiagnostic}
\end{eqnarray}}

The amplitude $A$ of the large-scale condensate can be expressed in terms of the conserved energy $E$ of the initial condition:
\begin{eqnarray}
E=A^2 \la |\bnabla \psi_0|^2 \ra + \la |\bnabla \psi_\eta|^2 \ra \, ,
\end{eqnarray}
that is,
\begin{eqnarray}
A= \sqrt{\frac{E- \la |\bnabla \psi_\eta|^2 \ra}{\la |\bnabla \psi_0|^2 \ra}} =  \frac{1}{k_0} \sqrt{E- \la |\bnabla \psi_\eta|^2 \ra} \, . \label{tempA}
\end{eqnarray}
For monoscale topography the expression of $\la |\bnabla \psi_\eta|^2 \ra$ reads:
\begin{eqnarray}
\frac{\la |\bnabla \psi_\eta|^2 \ra}{E_\eta} = \frac{\tilde{k}_\eta^4}{(\tilde{k}_\eta^2-1)^2} \, ,
\end{eqnarray}
where $\tilde{k}_\eta=k_\eta/k_0$, and after substitution in (\ref{tempA}) one obtains the following expression for the dimensionless amplitude of the condensate $\tilde{A}=k_0 A/\sqrt{E_\eta}$:
\begin{eqnarray}
\tilde{A}= \sqrt{\tilde{E} -  \tilde{E}_c} \, , \qquad \text{with } \quad \tilde{E}_c = \frac{\tilde{k}_\eta^4}{(\tilde{k}_\eta^2-1)^2} \, . \label{amplAconservative}
\end{eqnarray}
This prediction takes a particularly simple form in the limit of scale separation between the domain size and the wavelength of the topography, $\tilde{k}_\eta \gg 1$, which leads to $\tilde{E}_c \simeq 1$:
\begin{eqnarray}
\tilde{A}= \sqrt{\tilde{E} -  1} \, . \label{amplAconservativelimit}
\end{eqnarray}
As illustrated by the plot in figure~\ref{fig:AvsE}, equation (\ref{amplAconservativelimit}) displays a transition to large-scale condensation above a critical value $E_\eta$ of the energy of the initial condition. An interpretation of this result is that the topography plays the role of a reservoir that can host a small-scale flow of energy up to $E_\eta$. Any energy beyond $E_\eta$ spills out of this reservoir and is subject to inverse energy transfers. This excess energy eventually ends up accumulating in a large-scale condensate, hence the nonzero value of $A$.

The enstrophy associated with the condensed branch is:
\begin{eqnarray}
\frac{Q}{\la \eta^2\ra}=\frac{1}{\tilde{k}_\eta^2} \left(\tilde{E} - 1 - \frac{1}{\tilde{k}_\eta^2-1} \right) \, ,
\end{eqnarray}
valid only when $A$ exists, that is for $\tilde{E}>\tilde{E}_c$. In the scale separation limit this expression reduces to:
\begin{eqnarray}
\frac{Q}{\la \eta^2\ra}=\frac{1}{\tilde{k}_\eta^2} \left(\tilde{E} - 1 \right) \, , \label{Qcondensate}
\end{eqnarray}
valid for $\tilde{E}>1$.

\begin{figure}
    \centerline{\includegraphics[width=8 cm]{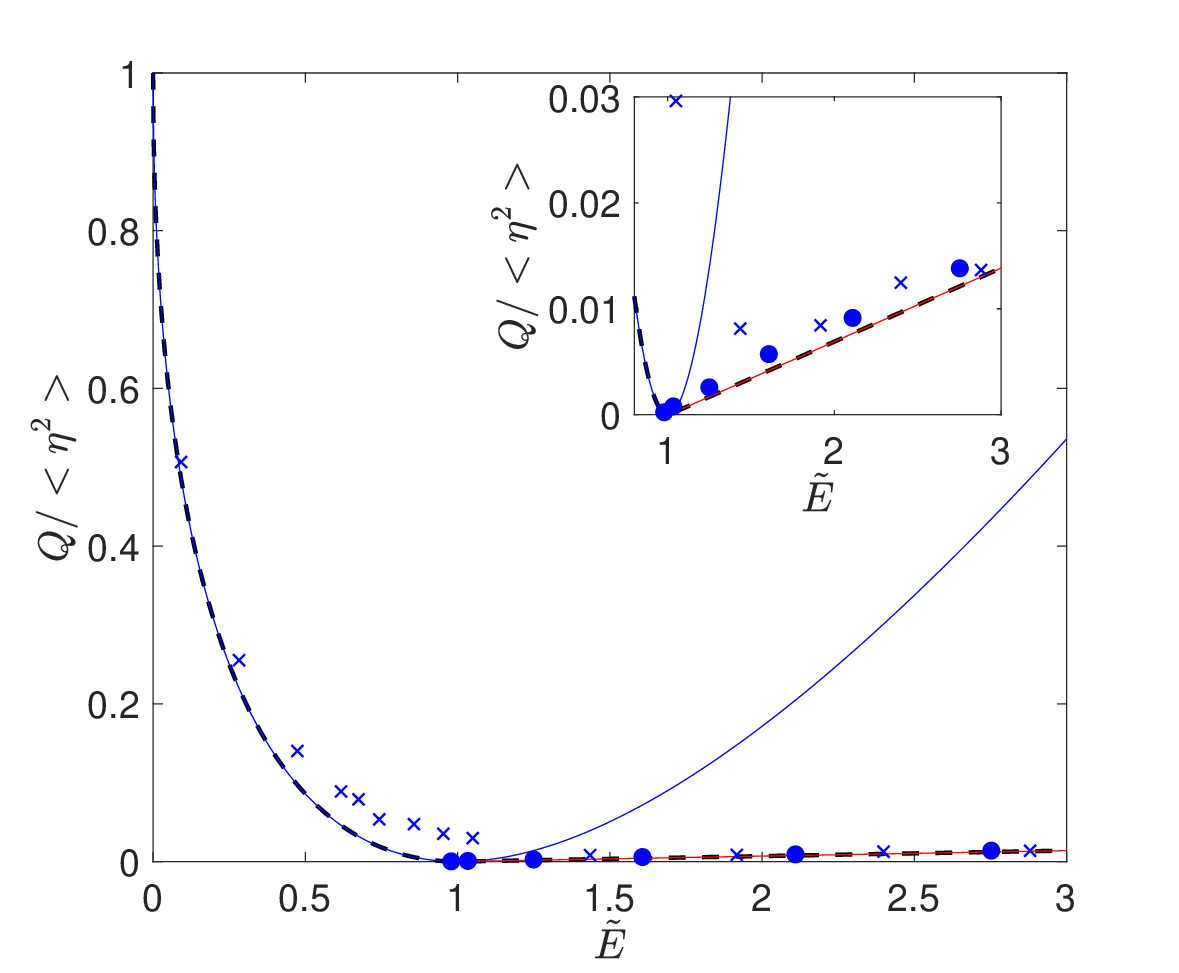}  }
   \caption{Enstrophy $Q$ in the (quasi-)stationary state as a function of the (conserved) energy of the initial condition $\tilde{E}$, using the same symbols as in figure~\ref{fig:AvsE}. Also shown are the results of the minimization: BH branch in blue and condensed branch in red, for $\tilde{k}_\eta=12$. The absolute minimum (over the branches) is shown as a black dashed line. The DNS data is close to the BH branch for $\tilde{E} \leq 1$ and close to the condensed branch for $\tilde{E} \geq 1$. The inset is a zoom on the data for $\tilde{E} \geq 1$. The departure from the absolute minimum near the condensation threshold is due to the extra vortices pinned to the topography. \label{fig:QvsE}}
\end{figure}

In figure~\ref{fig:QvsE} we show the enstrophy $Q$ as a function of the energy $E$ for both the BH branch and the condensed branch. One can check that the enstrophies $Q(E)$ evaluated on the two branches (\ref{QBH}) and (\ref{Qcondensate}) are equal and tangent to one another at the critical value $\tilde{E}=\tilde{E}_c$. For $\tilde{E}<\tilde{E}_c$ the absolute enstrophy minimizer is the BH branch, while for $\tilde{E}>\tilde{E}_c$ the absolute minimizer corresponds to the condensed branch. Absolute minimization of enstrophy thus predicts a transition to condensation as $E$ exceeds $E_c$, the amplitude of the condensate being given by (\ref{amplAconservative}). These predictions are particularly compact in the limit of scale separation, where the dimensionless threshold for condensation reduces to $\tilde{E}_c=1$ and the amplitude of the condensate is given by (\ref{amplAconservativelimit}).


\section{Numerical simulations\label{sec:Numerical}}

\begin{figure}
    \centerline{\includegraphics[width=14 cm]{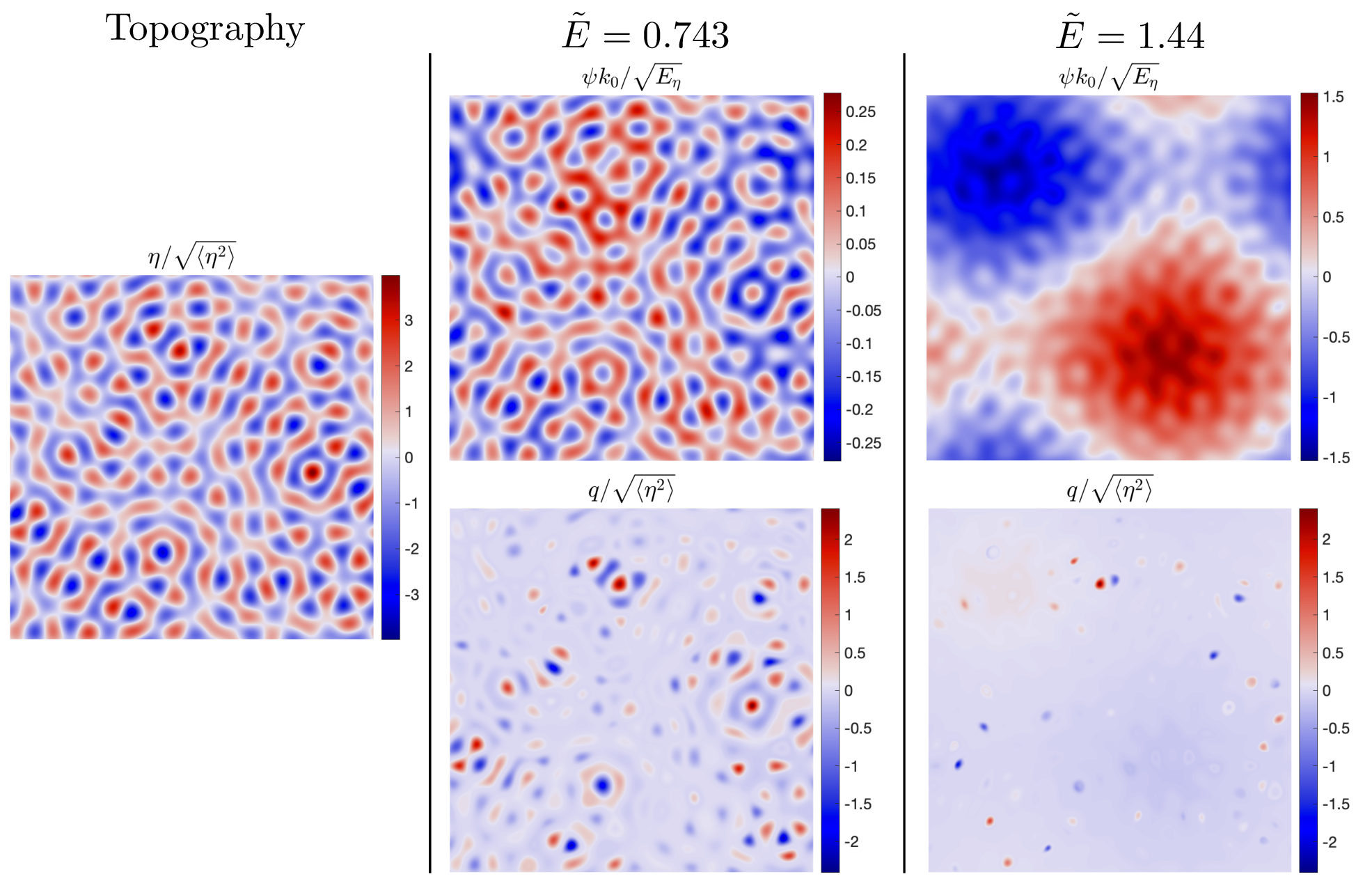} }
   \caption{For the topography shown on the left-hand panel, a numerical run initialized with energy $\tilde{E}=0.743$ displays no condensation (center column), whereas a run initialized with $\tilde{E}=1.44$ settles in a condensed state in the long-time limit (right-hand column). \label{fig:snapshots}}
\end{figure}

To test the predictions above, we perform DNS of equations (\ref{eqq}-\ref{defq}) by time-stepping $\psi$ using a pseudo-spectral solver. The resolution is $1024^2$ before de-aliasing following the $2/3$ rule. We first consider the enstrophy-decaying energy-conserving case, ${\cal F}=0$ and $\kappa=0$ with a small value of the hyperviscosity $\nu k_0^3/ \sqrt{E_\eta}=2\times 10^{-10}$. 
The topography is approximately monoscale and centered around the dimensionless wavenumber $\tilde{k}_\eta=12$. Specifically, we pick random values for the Fourier coefficients of all the modes such that $12-1/4 \leq kL \leq 12+1/4$. All the numerical runs are performed using the same realization of the topography, displayed in the left-hand panel of figure~\ref{fig:snapshots}.

We consider two types of initial conditions:
\begin{itemize}
\item The `$\psi$ initial condition': we set some initial complex Fourier amplitude at wavenumber $(k_x L,k_y L)=(0,2)$ for $\psi$ to match the target initial value of $\tilde{E}$. 
\item The `$\phi$ initial condition': we set some initial complex Fourier amplitude at wavenumber $(k_x L,k_y L)=(0,2)$ for a field $\phi(x,y)$, before initializing $\psi$ as $\psi(x,y,t=0)=\psi_\eta(x,y)+\phi(x,y)$.
\end{itemize}
Both types of initial conditions correspond to initial perturbations at rather large scale. Starting from the $\psi$ initial condition, the flow spontaneously develops smaller structures through interaction with the topography. By contrast, the $\phi$ initial condition readily includes the small-scale flow structure predicted by enstrophy minimization. This initial condition only allows to achieve an initial energy $\tilde{E}\geq 1$. By contrast, one can achieve any positive initial value for the energy $\tilde{E}$ using the $\psi$ initial condition.


As discussed by SY, in such simulations the enstrophy rapidly decays from its initial value. By contrast, the energy decreases extremely slowly and is conserved to a very good approximation. We wish to characterize the (quasi-)stationary state obtained once the transient has subsided. In figure~\ref{fig:snapshots} we show snapshots obtained at the end time of two simulations differing in the value of the initial energy. The streamfunction $\psi$ agrees qualitatively with the theory: the simulation with $\tilde{E}<1$ displays no large-scale condensation, whereas the simulation with  $\tilde{E}>1$ displays a strong large-scale condensate. By contrast, in line with previous studies the PV $q$ exhibits strong isolated vortices that are not predicted by enstrophy minimization~\citep{zhao2019influence,siegelman2023two,lacasce2024vortices}. When condensation arises, however, these vortices appear to be rather sparsely distributed above the faint background PV structure of the condensed branch, given by~(\ref{MEH}) with $\mu=-k_0^2$. There is thus hope for large-scale condensation to be quantitatively predicted by enstrophy minimization despite the emergence of the isolated vortices.

To characterize the condensation transition quantitatively, we extract from the DNS the amplitude $A$ \cor{as defined in equation (\ref{eq:Adiagnostic})}. In figure~\ref{fig:AvsE} we show the resulting dimensionless condensate amplitude $\tilde{A}$ as a function of the dimensionless initial energy $\tilde{E}$. The numerical data are in very good agreement with the theoretical prediction from enstrophy minimization. The $\phi$ initial condition leads to particularly good agreement because this initial condition is close to the enstrophy minimizer, thus preventing the emergence of strong isolated PV vortices. The more generic $\psi$ initial condition also leads to very good quantitative agreement. There is some departure from the theoretical prediction in the immediate vicinity of the bifurcation threshold only, as a consequence of the isolated vortices pinned to the topography.

In figure~\ref{fig:QvsE}, we plot the equilibrated enstrophy $Q$ extracted from the DNS. Above the transition $Q$ is significantly lower than the enstrophy of the BH branch, because the latter is not the absolute minimizer. The equilibrated $Q$ lies extremely close to the prediction of the condensed branch for DNS initialized using the $\phi$ initial condition. It is slightly greater than (but close to) this absolute minimum for the $\psi$ initial condition, as a result of the extra isolated vortices. Once again, this slight departure from the absolute minimum of $Q$ has only a modest impact on the value of the condensate amplitude $A$, which closely follows the theoretical prediction.

\section{Forced-dissipative case\label{sec:Forced}}

The natural objection to the variational approaches to computing the statistically steady states of fluid flows is that these approaches are usually based on conservative dynamics, while real flows are forced and dissipative. For instance, the statistical mechanics of 2D flows assumes that energy and enstrophy are conserved and keep their initial values, while DNS point to a transient phase with a robust decrease in enstrophy. In that respect, selective decay (enstrophy minimization) is one step closer to realistic flows, allowing for such a net decrease in enstrophy as the system evolves towards the minimizer, in line with the forward enstrophy cascade of 2D flows. However, selective decay also assumes that energy is conserved and set by the initial condition, whereas natural and laboratory flows typically receive energy from some external forcing and dissipate it through a combination of damping mechanisms. To make contact between such forced-dissipative systems and the variational framework discussed above, we now consider equation (\ref{eqq}) with nonzero forcing ${\cal F}$ injecting energy into the system, and nonzero friction coefficient $\kappa$ extracting energy from the system (we also retain the hyperviscous coefficient). The goal is at a fundamental level: more than describing realistic flows of geophysical interest, we wish to establish a direct quantitative connection between the variational approach described above and forced-dissipative systems. We thus illustrate how weak forcing and dissipation select a unique state out of the continuum of solutions to the energy conserving problem predicted by selective decay.

\subsection{Selection of minimum-enstrophy states}

\begin{figure}
    \centerline{\includegraphics[width=8 cm]{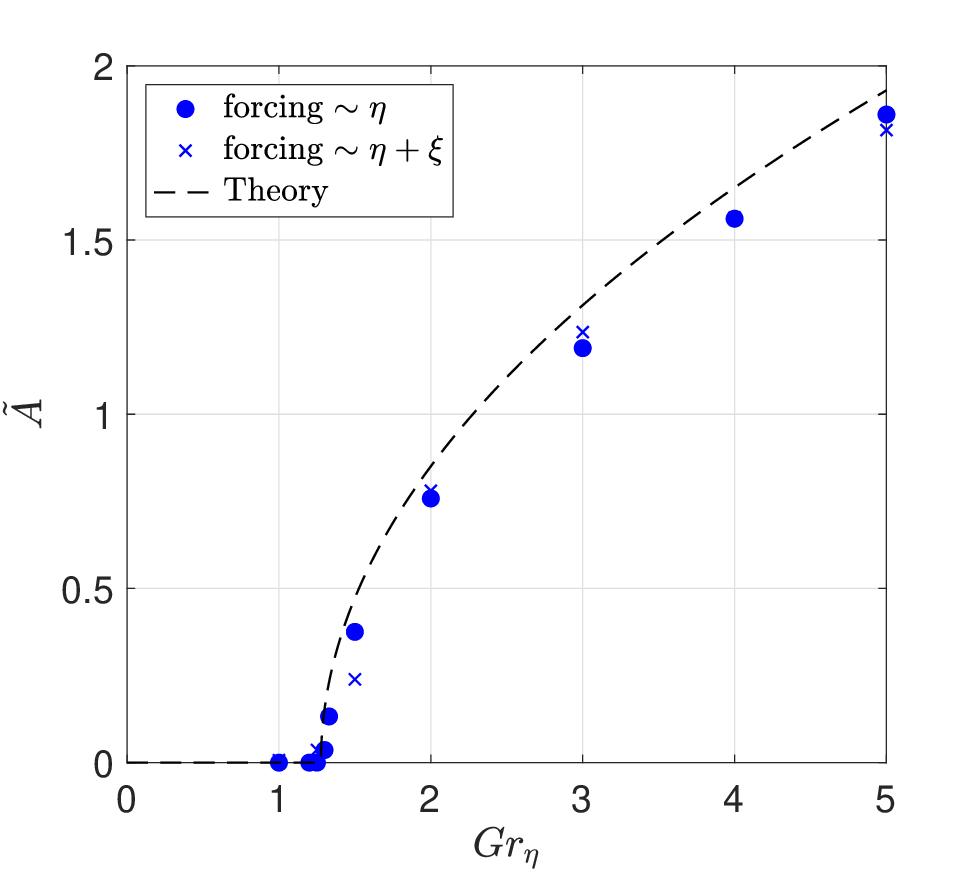}} 
   \caption{Condensate amplitude as a function of the forcing expressed in terms of the Grashof number. The DNS data points are in good agreement with the theoretical prediction  (\ref{eq:thforcing}) associated with the selection of minimum-enstrophy solutions (dashed line).\label{fig:AvsGr}}
\end{figure}

Consider equation~(\ref{eqq}) with weak forcing and damping: 
\begin{eqnarray}
& & \partial_t q + J(\psi,q) =\epsilon[ {\cal F}^{(1)}(x,y)+{\cal D}^{(1)} ] \, , \label{eqqexp}\\
& & q= \Delta \psi + \eta 
\end{eqnarray}
where $\epsilon \ll 1$ is a small parameter, ${\cal D}^{(1)}=-\kappa^{(1)} \Delta \psi - \nu^{(1)} \Delta^3 \psi$, and the topography $\eta$ is ${\cal O}(1)$. Expand the fields in powers of~$\epsilon$:
\begin{eqnarray}
q(x,y,t) & = & q^{(0)}(x,y,T)+\epsilon q^{(1)}(x,y,T) + \dots \, , \\
\psi(x,y,t) & = & \psi^{(0)}(x,y,T)+\epsilon \psi^{(1)}(x,y,T) + \dots \, ,
\end{eqnarray}
where we have introduced the slow time variable $T=\epsilon t$, that is, $\partial_t \to \partial_t +\epsilon \partial_T$.

Collecting terms of order ${\cal O}(\epsilon^0)$ leads to the set of equations discussed in the preceding section:
\begin{eqnarray}
& & \partial_t q^{(0)} + J(\psi^{(0)},q^{(0)}) = 0 \, , \qquad  q^{(0)}= \Delta \psi^{(0)} + \eta \, .
\end{eqnarray}
We consider a solution to these equations that lies on the condensed branch, allowing the amplitude of the condensate to vary slowly in time:
\begin{eqnarray}
\psi^{(0)}(x,y,T)  & = &  \psi_\eta(x,y) + A(T) \, \psi_0(x,y) \, .  \label{ansatzcond}
\end{eqnarray}
Following the traditional approach, the slow evolution equation for the amplitude $A(T)$ can be obtained from a solvability condition at next order. An alternate and more enlightening way of obtaining this amplitude equation consists of writing the energy evolution equation to lowest order. Upon multiplying equation (\ref{eqqexp}) with $\psi$ before averaging over $x$ and $y$ and neglecting terms of order $\epsilon^2$, one obtains:
\begin{eqnarray}
\frac{\mathrm{d}}{\mathrm{d}T} \left( \frac{\tilde{A}^2}{2} \right) & \simeq & -\frac{ \left< {\cal F}^{(1)} \eta \right>}{\la \eta^2 \ra} -  \kappa^{(1)} - \nu^{(1)} k_\eta^4  -\frac{\left< {\cal F}^{(1)} \psi_0 \right>}{k_0 \sqrt{E_\eta}}  \tilde{A} - \kappa^{(1)} \tilde{A}^2 \, , \label{eqAtemp}
\end{eqnarray}
where, for brevity, we restrict attention to the scale separation limit $\tilde{k}_\eta \gg 1$.
After dividing by~$\kappa^{(1)}$, equation (\ref{eqAtemp}) can be recast in terms of the original (non-expanded) variables as:
\begin{eqnarray}
\frac{1}{2 \kappa}\frac{\mathrm{d}}{\mathrm{d}t} \tilde{A}^2 & = & Gr_\eta-1-\frac{\nu k_\eta^4}{\kappa} - \tilde{A}^2 +Gr_0 \tilde{A} \, . \label{eqAgeneral}
\end{eqnarray}
\cor{For body-forced flows the dimensionless strength of the forcing is usually quantified by a Grashof number~\citep{doering1995applied}. In equation (\ref{eqAgeneral}) we have thus introduced a} topographic-scale Grashof number $Gr_\eta$ and a domain-scale Grashof number $Gr_0$, defined respectively as:
\begin{eqnarray}
Gr_\eta  =  - \frac{\left< {\cal F} \eta\right>}{\kappa \left< \eta^2 \right>} \, \qquad \text{and} \qquad Gr_0  =  - \frac{\la{\cal F} \psi_0 \ra }{k_0 \kappa \sqrt{E_\eta}} \, .
\end{eqnarray}
The forcing thus enters the evolution equation~(\ref{eqAgeneral}) for the amplitude of the condensate in two different ways: through its correlation with the smaller-scale topography and through its correlation with the domain-scale condensate. In the following we consider these two situations independently.

\subsection{Topographic-scale forcing: continuous transition to condensation}

Consider first the situation where there is no forcing at large-scale, $\la{\cal F} \psi_0 \ra =0$ and therefore $Gr_0=0$. Equation (\ref{eqAgeneral}) reduces to:
\begin{eqnarray}
\frac{1}{2 \kappa}\frac{\mathrm{d}}{\mathrm{d}t} \tilde{A}^2 & = & Gr_\eta-Gr_\eta^{(c)}- \tilde{A}^2  \, ,  \label{eqAsmallscale}
\end{eqnarray}
where we have introduced a critical Grashof number $Gr_\eta^{(c)}=1+{\nu k_\eta^4}/{\kappa}$.

For $Gr_\eta<Gr_\eta^{(c)}$, $\tilde{A}^2$ decreases and becomes negative according to this equation. The ansatz (\ref{ansatzcond}) is then invalid and instead the system ends up on the non-condensed BH branch. By contrast, a steady solution $\tilde{A} \geq 0$ exists for $Gr_\eta \geq Gr_\eta^{(c)}$. We thus predict the following value for the equilibrated amplitude of the condensate:
\begin{eqnarray}
\tilde{A} &= & 0 \, \qquad \text{for } Gr_\eta<Gr_\eta^{(c)} \, , \\
\tilde{A} &= & \sqrt{Gr_\eta-Gr_\eta^{(c)}}  \, \qquad \text{for } Gr_\eta \geq Gr_\eta^{(c)} \, . \label{eq:thforcing}
\end{eqnarray}
This prediction corresponds to a supercritical transition to condensation as the dimensionless forcing strength $Gr_\eta$ increases. The bifurcation threshold reduces to the compact expression $Gr_\eta^{(c)} = 1$ in the limit where hyperviscosity damps energy at a negligible rate as compared to friction, $\nu k_\eta^4\ll \kappa$.



As discussed above, the forcing enters the amplitude equation~(\ref{eqAsmallscale}) through its correlation with the topography. We first illustrate this situation numerically using a forcing that is perfectly correlated with the topography:
\cor{\begin{eqnarray}
 {\cal F}(x,y)=- Gr_\eta \, \kappa \, \eta(x,y) \, .
 \end{eqnarray}}
 We consider the same realization of the topography as in the preceding sections, together with damping coefficients $\kappa /(k_0 \sqrt{E_\eta})=0.0021$ and $\nu k_0^3/ \sqrt{E_\eta}=2.8 \times 10^{-8}$. The corresponding critical Grashof number is $Gr_\eta^{(c)} \simeq 1.28$. We perform a suite of numerical simulations for various values of $Gr_\eta$, extracting the amplitude of the condensate once the transient has subsided. We show $\tilde{A}$ as a function of $Gr_\eta$ in figure~\ref{fig:AvsGr}, together with the theoretical prediction~(\ref{eq:thforcing}). The agreement is very good, both for the threshold value $Gr_\eta^{(c)}$ and for the amplitude of the condensate above threshold. 
Together with the simple forcing described above, we have considered a second forcing structure where ${\cal F}(x,y)$ is proportional to $\eta(x,y)+\xi(x,y)$, where $\xi(x,y)$ is a random field with the same statistics as $\eta(x,y)$ but corresponding to another realization (the realizations of $\eta$ and $\xi$ are the same for the entire suite of simulations). Even though the resulting forcing ${\cal F}(x,y)$ is less correlated to the topography than in the previous suite of simulations, the condensate amplitude is again well predicted by equation~(\ref{eq:thforcing}), as shown in figure~\ref{fig:AvsGr}.

\subsection{Large-scale forcing: discontinuous transition to condensation}

Consider now the situation where the forcing entering the amplitude equation~(\ref{eqAgeneral}) is at large scale only, $Gr_\eta=0$ and $Gr_0\neq 0$. For simplicity, we also consider that the hyperviscous contribution to this amplitude equation is negligible, $\nu k_\eta^4 \ll \kappa$. Equation~(\ref{eqAgeneral}) reduces to:
\begin{eqnarray}
\frac{1}{2 \kappa}\frac{\mathrm{d}}{\mathrm{d}t} \tilde{A}^2 & = & Gr_0 \tilde{A}  -1 - \tilde{A}^2 \, , \label{eqALSF}
\end{eqnarray}
with steady solutions:
\begin{equation}
\tilde{A}_\pm = \frac{1}{2} \left( Gr_0 \pm \sqrt{Gr_0^2-4} \right) \, . \label{eq:subcritical_prediction}
\end{equation}
These solutions exist only for $Gr_0 \geq 2$, while for $Gr_0<2$ the system lies on the uncondensed BH branch, with $\tilde{A}=0$. One can furthermore check that $A_-$ is unstable according to equation (\ref{eqALSF}), while $A_+$ is a stable branch of solutions. We thus predict a subcritical bifurcation to large-scale condensation as $Gr_0$ exceeds the threshold value $Gr_0^{(c)}=2$. It is interesting to notice that the same supercritical bifurcation of the energy-conserving system induces either a supercritical or a subcritical bifurcation in the weakly forced-dissipative system, depending on the structure of the forcing.

\begin{figure}
    \centerline{\includegraphics[width=8 cm]{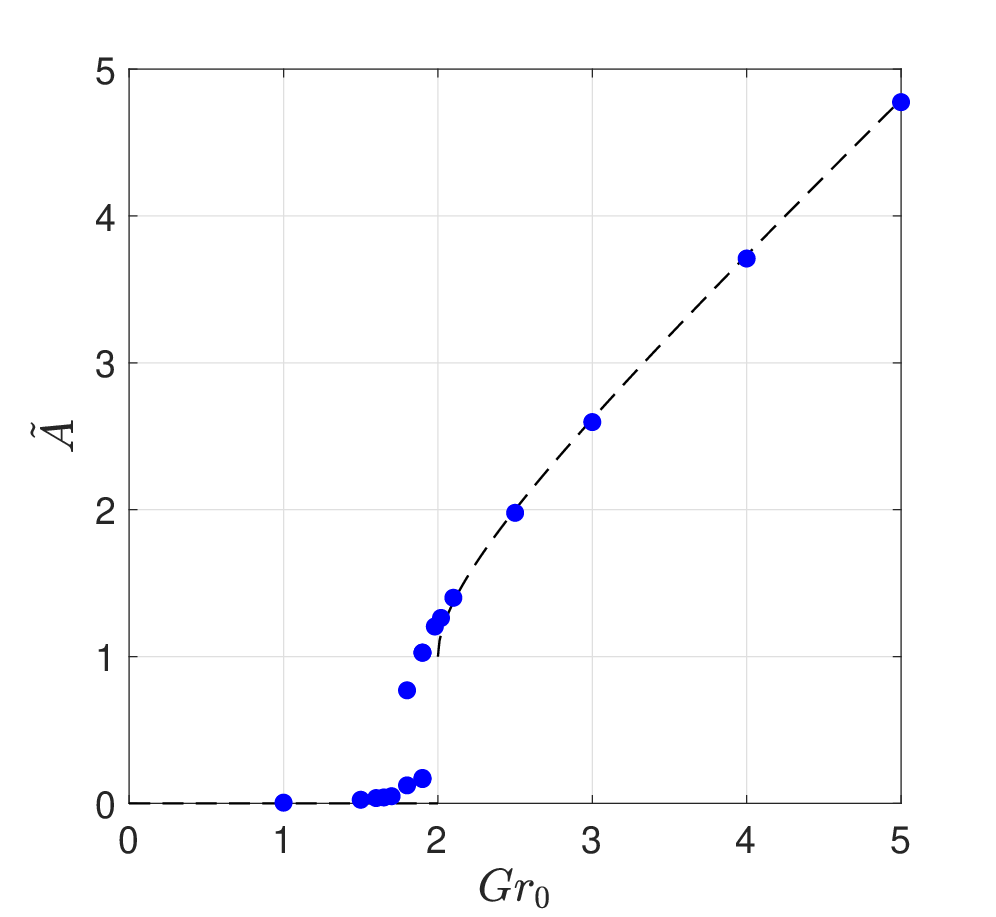}} 
   \caption{Condensate amplitude as a function of the large-scale forcing amplitude expressed in terms of $Gr_0$. The DNS data points are in good agreement with the theoretical prediction $A_+$ in (\ref{eq:subcritical_prediction}) associated with the selection of minimum-enstrophy solutions (dashed line).\label{fig:AvsGr0_LSF}}
\end{figure}

We investigated this situation numerically with the same resolution and the same realization of the topography as before. The friction coefficient is still $\kappa/(k_0 \sqrt{E_\eta})=0.0021$, while we employ the much smaller value of the hyperviscosity $\nu k_0^3/ \sqrt{E_\eta}=2\times 10^{-10}$, which ensures that $\nu k_\eta^4 \ll \kappa$. The forcing is:
\cor{\begin{equation}
{\cal F}(y)=-\sqrt{2 E_\eta} k_0 \kappa \, Gr_0 \, \cos(y) \, , \label{eq:shapeforcing}
\end{equation}}
associated with a structure $\psi_0=\sqrt{2}\cos y$ for the large-scale flow. To investigate possible bistability, the initial condition is either very low amplitude fluctuations or the end state of a previous simulation.

The amplitude of the condensate is shown as a function of $Gr_0$ in figure~\ref{fig:AvsGr0_LSF}. Once again, the agreement with the theory is excellent at finite distance from the instability threshold, with slight discrepancies near threshold. In particular, the amplitude of the condensate is very well captured by the upper-branch in (\ref{eq:subcritical_prediction}) for $Gr_0 \geq 2$. We observe only a very narrow region of bistability, possibly because the uncondensed branch has other directions of instability, besides direct condensation following~(\ref{eqALSF}). Additionally, the region of stability of the condensed branch extends somewhat below $Gr_0=2$, down to $Gr_0\simeq 1.8$. This $10\%$ correction in threshold value may be a consequence of the extra vortices pinned to the topography. As for the previous cases considered above, we conclude that the amplitude of the condensate is accurately predicted by the theory, except in the immediate vicinity of the instability threshold where the pinned vortices impact the dynamics of the system.

\cor{\subsection{Structure of the condensate}
We have mainly discussed the amplitude of the condensate, setting aside its degenerate structure $\psi_0(x,y)$. The latter depends on the relative magnitudes of the complex Fourier coefficients $\hat{\psi}_{(1/L,0)}$ and $\hat{\psi}_{(0,1/L)}$, as well as on their phases. In simulations of the energy-conserving system, the precise structure of the condensate (see e.g. figure~\ref{fig:snapshots}) seems to be determined predominantly by the precise realization of the topography and, to a lesser extent, by the initial condition. When driving the system with the domain-scale forcing~(\ref{eq:shapeforcing}) together with moderately small friction, the condensate always adopts the large-scale structure of the forcing, $\psi_0(x,y)=\sqrt{2} \cos(y)$, and the theoretical predictions are accurately satisfied. However, if $\epsilon$ on the right-hand side of equation~(\ref{eqqexp}) is made extremely small (very weak forcing and dissipation), there seems to be a competition between the forcing, which favors the structure $\psi_0(x,y)=\sqrt{2} \cos(y)$, and the precise realization of the topography, which favors an alternate structure, such as the one visible in figure~\ref{fig:snapshots}. The prediction (\ref{eq:subcritical_prediction}) for the condensate amplitude does not hold if the system adopts this alternate condensate structure. An in-depth characterization of such phase dynamics in the very-weak-drag regime would deserve a dedicated study that goes beyond the scope of the present article. One way to lift the degeneracy of the condensate structure consists in including weak domain-scale topography in a perturbative fashion, as described in appendix~\ref{sec:domaintopo}. Then a single condensate structure $\psi_0(x,y)$ proportional to the domain-scale topography arises, the condensate amplitude being still determined by forcing and dissipation (or by the initial energy for the energy-conserving system). }

\section{Conclusion\label{sec:Conclusion}}

We have characterized the transition to large-scale condensation for 2D flows over smaller-scale topography, revisiting the selective decay or minimum-enstrophy hypothesis of BH. We have computed a condensed branch of solutions to the minimization problem. This branch of solutions is the absolute enstrophy minimizer above some threshold value of the (conserved) energy. As this control parameter increases, selective decay predicts a continuous transition to large-scale condensation. This bifurcation adds to the various `phase transitions' of quasi-2D fluid systems predicted based on the statistical mechanics of 2D flows, such as the transition in flow structure as the aspect ratio of a 2D rectangular fluid domain is varied~\citep{chavanis1996classification,bouchet2012statistical}. The present bifurcation is also very similar to the emergence of a Rossby-wave in the enstrophy-minimizing state of a channel flow on the $\beta$ plane~\citep{young1987selective}, the connection between the two problems being apparent if $\beta$ is interpreted as a uniform topographic slope.

For simplicity, we have illustrated the present theory using monoscale topography, but the approach is easily extended to more complex topography. As compared to the recent study of flows above topography by SY, the main qualitative difference is probably that SY include large topographic variations at the domain scale. \cor{SY report a transition from isolated pinned vortices to freely roaming vortices as the initial energy exceeds a threshold value $E_\sharp=\la | \bnabla \Delta^{-1} \{ \eta \} |^2 \ra$, omitting the prefactor $1/2$ for consistency with our definition for $E$ (SY identify $E_\sharp$ as the energy of a flow with homogeneous PV $q=0$, which for monoscale topography reduces to the energy scale $E_\eta$ that we introduced on dimensional grounds).} By contrast, in the absence of topography at the domain scale we observe a transition to condensation as the initial energy exceeds a threshold value $E_c=\la |\bnabla \psi_\eta|^2 \ra$. Interestingly, \cor{while $E_c \neq E_\sharp$ for arbitrary topography,} $E_c$ and $E_\sharp$ reduce to the same expression in the limit of separation between the topographic scale and the domain scale. \cor{This indicates} that SY's study and the present study may correspond to two manifestations of the same transition. An appealing aspect of the present setup is that selective decay provides quantitative predictions for the amplitude of the condensate above threshold.

Indeed, another goal of the present study is to assess the predictive skill of such a variational approach through comparison with DNS.
DNS of the energy-conserving system show that selective decay quantitatively predicts the amplitude of the condensate. For initial conditions tailored to be close to the absolute minimizer, the amplitude of the condensate is in excellent agreement with the theory. For more generic initial conditions isolated vortices arise, polluting the potential vorticity field $q$. The vortices affect the very near-threshold behavior of the condensate amplitude only, the agreement remaining very good at larger distance from threshold.
That the vortices predominantly affect the near-threshold behavior probably originates from the fact that they become more scarce as condensation proceeds. Indeed, the isolated vortices are located at the local extrema of the total streamfunction, as discussed in SY. In the absence of large-scale condensation, the streamfunction resembles the topography, with many such extrema. By contrast, at finite distance above threshold the condensate dominates the total streamfunction, and the latter has fewer local extrema to host isolated vortices (see figure~\ref{fig:snapshots}). 


The above-mentioned dependence on initial conditions is perhaps a complication of the energy-conserving system. After all, natural and laboratory flows are typically forced and dissipative, with energy supplied by an external source and removed by some combination of damping mechanisms. The energy of the equilibrated state is then an emergent property of the system, as opposed to the initial energy of the system being a control parameter of the variational approaches to conservative flows. As such, the variational approaches to energy-conserving systems cannot predict the strength of forced-dissipative flows.
With the goal of bridging the gap between these two situations, we have introduced a perturbative expansion that illustrates how weak forcing and dissipation select the equilibrated state of the flow out of the continuum of states predicted by selective decay (or by the statistical mechanics of 2D flows). \cor{We have restricted attention to steady forcing for simplicity, but the approach probably holds for slowly varying forcing.}

This selection mechanism is very similar to -- and was inspired by -- the selection of solitary wave solutions to the complex Ginzburg-Landau equation by weak dissipation~\citep{fauve1990solitary}.
The theoretical predictions for the resulting amplitude of the condensate agree quantitatively with DNS of the forced-dissipative system. Interestingly, the same continuous transition to condensation for the energy-conserving system turns into a discontinuous transition when the system is forced at large scale, while it remains continuous when the system is forced at the smaller topographic scale. 

This method provides an avenue for predicting the equilibrated state of forced-dissipative flows based on variational approaches initially designed for conservative systems (selective decay or statistical mechanics). Its robustness and predictive skill remain to be further assessed through application to various systems of interest, be they numerical simulations, laboratory experiments or observational data.

\medskip
\noindent {\textbf{Acknowledgements.} {I would like to thank W.R. Young and A. Venaille for insightful discussions, and S. Fauve for raising my interest in the selection of conservative solutions in weakly forced dissipative systems.}

\medskip
\noindent {\textbf{Funding.} This research is supported by the European Research Council under grant agreement FLAVE 757239.}

\medskip
\noindent {\textbf{Declaration of interests.} The author reports no conflict of interest.}

\medskip
\noindent \textbf{Author ORCID.}  https://orcid.org/0000-0002-4366-3889.

\appendix

\section{Including weak domain-scale topography \label{sec:domaintopo}}

\cor{Weak domain-scale topography can be included through the following Poincaré-Lindstedt expansion:
\begin{eqnarray}
\eta(x,y) & = &\eta^{(0)}(x,y)+\epsilon \eta^{(1)}(x,y) \, , \\
\psi(x,y,t) & = & \psi^{(0)}(x,y)+\epsilon \psi^{(1)}(x,y) + \dots \, , \\
\mu & = & -k_0^2 +\epsilon \mu^{(1)} \, ,
\end{eqnarray}
where $\eta^{(0)}(x,y)$ denotes the monoscale topography adopted throughout the main text, $\eta^{(1)}(x,y)$ denotes domain-scale topography on the gravest modes ${\bf k}=(1/L,0)$ and ${\bf k}=(0,1/L)$, and we have expanded the Lagrange multiplier $\mu$ around its value $ -k_0^2 $ on the condensed branch. To order ${\cal O}(1)$, equation (\ref{MEH}) yields $\Delta \psi^{(0)} + k_0^2 \psi^{(0)} = - \eta^{(0)}$, with solution $\psi^{(0)}=A \psi_0(x,y)+\psi_\eta(x,y)$. To order ${\cal O}(\epsilon)$ equation (\ref{MEH}) yields $\Delta \psi^{(1)} + k_0^2 \psi^{(1)} =  \mu^{(1)} \psi^{(0)} - \eta^{(1)}$. The solvability condition is obtained by demanding that the right-hand side of this equation have no projection onto the gravest Fourier modes, which simply yields $\mu^{(1)} A \psi_0 = \eta^{(1)}$. Using $\la \psi_0^2 \ra=1$ to eliminate $\mu^{(1)}$, we obtain the following expression for the structure of the condensate: 
\begin{eqnarray}
\psi_0(x,y) & = & \pm \frac{\eta^{(1)}(x,y)}{\sqrt{\la (\eta^{(1)})^2\ra}} \, .
\end{eqnarray}
Weak domain-scale topography thus lifts the degeneracy of the condensate structure. The amplitude $A$ of the condensate is still determined following the procedures described in the main text (initial energy for the energy-conserving system ; forcing and damping for the forced-dissipative system).}



\bibliographystyle{jfm}

\bibliography{topoQG}

\end{document}